\documentclass[aps,12 pt]{revtex4}
\usepackage{amssymb}
\usepackage{amsmath}
\usepackage{graphicx}
\usepackage{epsfig,amsmath}
\usepackage[bookmarksnumbered,linktocpage,pdfstartview=FitH]{hyperref}

\begin{document}

\title{Probing of quantum turbulence with radiating vortex loops}
\author{Sergey K. Nemirovskii\thanks{%
email address: nemir@itp.nsc.ru}}
\affiliation{Institute of Thermophysics, Lavrentyev ave, 1, 630090, Novosibirsk, Russia\\
and Novosibirsk State University, Novosibirsk}
\date{\today }

\begin{abstract}
The statistics of vortex loops emitted from the domain with quantum
turbulence is studied. The investigation is performed on the supposition
that the vortex loops have the Brownian or random walking structure with the
generalized Wiener distribution. The main goal is to relate the properties
of the emitted vortex loops with the parameters of quantum turbulence. The
motivation of this work connected with recent studies, both numerical and
experimental, on study of emitted vortex loops. This technique opens up new
opportunities to probe superfluid turbulence. We demonstrated how the
statistics of emitted loops is expressed in terms of the vortex tangle
parameters and performed the comparison with numerical simulations.\newline
PACS number(s): 67.25.dk, 47.37.+q
\end{abstract}

\maketitle

\emph{Scientific Background and motivations.} Quantum turbulence (QT) in
superfluids is one of the most fascinating phenomena in the theory of
quantum fluids \cite{Nemirovskii2013}. In general, the vortex tangle
composing QT, consists of a set of vortex loops of different lengths and
having a random structure. Question of an arrangement of the vortex tangle
is a key problem of the theory of QT. Recently, a number of works devoted to
the study of the vortex tangle structure with the use of emitted from the
turbulent domain have been performed \cite{Nakatsuji2014},\cite{Nago2013},%
\cite{Kondaurova2012},\cite{Walmsley2013}. This technique opens up new
opportunities for research QT.

One of the main problem in this activity is to relate the properties of the
emitted vortex loops with the parameters of quantum turbulence. In the
present work we propose an analytical approach that allows to relate the
statistics of vortex loops with the parameters of the real vortex tangle
This approach is based on the Gaussian model of the vortex tangle, which
describes the latter as a set of vortex loops having a random walking
structure with the generalized Wiener distribution,\cite{Nemirovskii2008}.
Then we apply our results for data processing of numerical work \cite%
{Nakatsuji2014} who studied quantum turbulence driven by an oscillating
sphere. Generation of quantum turbulence by oscillating objects is very
important topic in this field (see, e.g. \cite{Chagovets2009},\cite%
{Sheshin2008},\cite{Gritsenko2014})

\emph{Flux of vortex loops emitted from quantum turbulence.} In this
paragraph we very briefly describe main ideas leading to the theory of
emission of vortex loops, details can be found in paper by the author \cite%
{Nemirovskii2010}. Vortex loops composing the vortex tangle can move as a
whole with a drift velocity $V_{l}$ depending on their structure and their
length $l$. The flux of the line length, energy, momentum etc., executed by
the moving vortex loops takes place. In the case of inhomogeneous vortex
tangle the net flux $\mathbf{J}$ of the vortex length due to the gradient of
concentration of the vortex line density $\mathcal{L}(x,t)$ appears. The
situation here is exactly the same as in classical kinetic theory with the
difference being that the "carriers" are not the point particles but the
extended objects (vortex loops), which possess an infinite number of degrees
of freedom with very involved dynamics.

To develop the theory of the transport processes fulfilled by vortex loops
(in spirit of classical kinetic theory) we need to know the drift velocity $%
V_{l}$ and the free path $\lambda (l)$\ for the loop of size $l$. Referring
to the paper \cite{Nemirovskii2010} we write down here the following result.
The drift velocity $V_{l}$ \ and the free path $\lambda (l)$ for the loop of
size $l$ are
\begin{equation}
V_{l}=C_{v}\beta /\sqrt{l\xi _{0}},\ \ \ \lambda (l)=1/2lb_{m}\mathcal{L}.
\label{Vl}
\end{equation}%
Quantity $\beta $ is $(\kappa /4\pi )\ln (\mathcal{L}^{-1/2}/a_{0})$, where $%
\kappa $ is the quantum of circulation and $a_{0}$ is the core radius, $%
C_{v} $ is numerical factor of the order of unity, $b_{m}$ is the numerical
factor , approximately equal to $\ b_{m}\approx 0.2$ . The $\xi _{0}$ is the
parameter of the generalized Wiener distribution, it is of order of the
interline space $\mathcal{L}^{-1/2}.$ The probability $P(x)$ for the loop of
length $l$ to fly the distance $x$ without collision is $P(x)=(1/\lambda
(l))\exp (-x/\lambda (l))$. Knowing the averaged velocity $V_{l}$ of loops,
and the probability $P(x)$ (both quantities are $l$ -dependent), we can
evaluate the spatial flux $\mathbf{J}$ of the vortex loops.

Let us consider the small area element placed at some point of the boundary
of domain containing QT and oriented perpendicularly to axis $x$ (See for
details Fig.2 of paper \cite{Nemirovskii2010}). The $x$ component of flux $%
\mathbf{J}$ of the number of loops executed by loops of sizes $l,$ placed in
$\theta ,\varphi $ direction, and remote from the area element at distance $%
R,$ can be written as:

\begin{equation}
J_{x}(l)=\frac{1}{4\pi }\int n(l,R,\theta ,\varphi )(V_{l}\cos \theta
)P(R)\sin \theta d\theta d\varphi dR.  \label{Jx(l)}
\end{equation}%
Here the quantity $(V_{l}\cos \theta )$ is just the $x$ component of the
drift velocity, the factor $P(R)dR$ is introduced to control an attenuation
of flux, due to collisions. In the spirit of classical kinetic theory, we
assume the local equilibrium is established.

In paper \cite{Nemirovskii2010} \ Eq. ( \ref{Jx(l)}) was the starting point
to develop the theory of diffusion of vortex loops in quantum turbulence,
therefore the density of loops $n(l)$\ was supposed to depend on spacial
position, in spherical coordinates $n=n(l,R,\theta ,\varphi )$). Here we put
another goal to study radiation of loops from the domain with uniform vortex
tangle. Supposing that inside domain the density of loops does not depend on
spatial coordinates ($n(l,R,\theta ,\varphi )=n(l))$, and integrating out
over solid angle $d\theta d\varphi $ and over position of loops $dR$ we
obtain the $x$-component of the loop flux through the domain boundary
\begin{equation}
J_{x}=\frac{1}{4}\int n(l)(\frac{\beta }{\sqrt{l\xi _{0}}})dl.
\label{J total}
\end{equation}

The flux $J_{x}$, described by formula \ref{J total} carries the loops of
different sizes in the normal to boundary direction and provides information
on the distribution of loops inside of the turbulent domain. However, since
the speed of loops depends on their sizes the initial distribution changes
as the vortex "cloud" propagates. For instance, in some time the small
vortex loops (practically rings) get ahead, and the measurements in a short
time will detect only small loops. That means that the experimental data
essentially depends on the position of detector and time time of
registration. In a stationary situation when the turbulence is maintained by
some means (counterflow or oscillating structures), the intensity of
detected loops should be steady in time.

\begin{figure}[h]
\includegraphics[width=12cm]{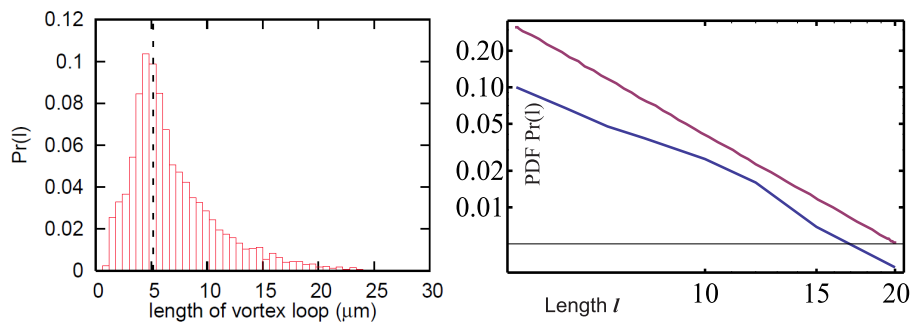}
\caption{(Color online) Left. The PDF of the length of emitted vortex loops
obtained in \protect\cite{Nakatsuji2014}. Right. Lower line is the PDF from
the left picture depicted in logarithmic coordinates. The upper curve is the
analytical PDF $\Pr_{an}(l)\approx 50l^{-3}$, obtained in this work. }
\label{PDF}
\end{figure}

\emph{Processing of experimental and numerical works.} As an illustration
let's treat the numerical work \cite{Nakatsuji2014} on the statistics of
emitted loops. Unfortunately, the pure experimental works are not too
reliable for a proper quantitative treatment. In work \cite{Nakatsuji2014}
the authors have studied the statistics of vortex loops emitted from quantum
turbulence driven by an oscillating sphere of radius $1\mu m$ with the
frequency $3$ kHz, and the amplitude $5.31$ $%
\mu
m$,. Counting of loops was made on the spherical boundary of radius 30 $\mu m
$ (centered at the center of the oscillating sphere). The result of this
counting is illustrated in the left graph in Fig.\ref{PDF} ,where the a
probability density function (PDF) $\Pr_{num}(l)$ of the length of the
emitted vortex loops is presented.

Let us try to consider this result from the position of the theory stated
above. Formula \ref{J total} describes the flux of loops per unit area. The
integrand in \ref{J total} is distribution of loops over their lengths in
the propagating loop "jet". The total number of emitted loops (per unit
time) is $N(l)=\int \mathbf{J}d\mathbf{S}$. The Gaussian model, used in our
consideration predicts that the density of loops $n(l)$ inside the domain
with vortex tangle is the power-like function $n(l)=Al^{-5/2}$. This
behaviour, however, has a low cutoff near the length $l\thicksim \xi _{0}$.
Below the cutoff there are a few loops of smaller sizes which do not
essentially affect the whole theory. From the graph for PDF $\Pr_{num}(l)$
in Fig. \ref{PDF} it is seen that the cutoff is $l\approx 5\ \mu m$. The
size $l\approx 5\ \mu m$ is the point of the maximum of the PDF $%
\Pr_{num}(l) $, where the power-like behaviour ceases. Using \ref{J total}
we get that the analytical PDF is $\Pr_{an}(l)=N(l)/\int N(l)dl\varpropto
l^{-3}$. Furthermore, taking into account that in the normalization factor $%
\int N(l)dl$, the value of integral is accumulated near the lower limit $%
l\thicksim 5\ \mu m$, we get that the analytical PDF approximatel is $%
\Pr_{an}(l)\approx 50l^{-3}$.
In the right picture of Fig \ref{PDF} we depicted PDF $\Pr_{num}(l)$
obtained in work \cite{Nakatsuji2014} and our $\Pr_{an}(l)\approx 50l^{-3}$
in the logarithmic coordinates. It can be seen that the power $-3$ is indeed
close to the numerical data, deviations appear near the low cutoff where the
Gaussian model does not work properly.\ As for absolute values of the PDFs
(numerical and analytical) they differ by a factor about $2\div 3$ (for
large loops). This can be explained by the fact that the vortex turbulence,
produced in numerical simulation is not the dense and uniform structure,
which was required in the analytical consideration. In fact, the interline
space obained from the data on total length (see Fig. 2 of paper \cite%
{Nakatsuji2014}) is about $20\mu m$ which is comparable with size of largest
loops. This implies that the VT is rather dilute.

\emph{Conclusion. }Summarizing we can conclude that the "clouds" of vortex
loops emitted from the turbulent superfluid helium provides an important
information on the structure of QT. However, due to very complicated
dynamics of the network of vortex loops, this information is hidden and
should be extracted with the use of appropriate formalism. It should be
understood that the described above procedure is greatly simplified, many
real features such as a possible anisotropy, the mutual friction, the
specific conditions of the turbulence generation have not been considered.
This is supposed to be performed in future.

The study was performed by grant from the Russian Science Foundation
(project N 14-29-00093) and by grant 13-08-00673 from RFBR
(Russian Foundation of Fundamental Research).


\end{document}